\documentclass{WileyMSP-template}
\usepackage{amsmath,amssymb,amsfonts}
\usepackage{tabularx} 
\usepackage{rotating}   
\usepackage{threeparttable}
\begin{document}

	\pagestyle{fancy}
	\rhead{\includegraphics[width=2.5cm]{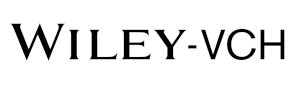}}

	\title{High efficiency controlled quantum secure direct communication with 4D qudits and Grover search algorithm}
	
	\maketitle

	
	\author{Ni-Shi Lu}
	\author{Ping Zhou*}
	
	\begin{affiliations}
		Ni-Shi Lu, Ping Zhou\\
		School of Physical and Electronic Sciences, Guangxi Minzu University,NanNing 530006,China\\
		Email Address:zhouping@gxmzu.edu.cn
		
	\end{affiliations}

	

	\begin{abstract}
		Currently, the progress of quantum secure direct communication (QSDC) is impeded by a fundamental trade off among control efficiency, security, and scalability. This study proposes an innovative controlled QSDC protocol based on a collaborative unitary sequence decoding paradigm to break this deadlock.Leveraging four dimensional single particle states, the protocol's core innovation lies in its three party decoding mechanism. The controller's authorization unlocks a specific unitary operation sequence, enabling the receiver to directly decode exclusively via quantum operations, eliminating the need for classical computational algorithms in conventional protocols. This tailored sequence underpins its high efficiency.The protocol also seamlessly incorporates decoy photon authentication, creating a multi layer defense against both external and internal attacks. Consequently, it achieves a remarkable qudit efficiency of 66.7\%, offering a significant performance improvement over existing schemes and an efficient, highly secure solution for future quantum networks.
	\end{abstract}
	

	\section{Introduction}
	The advent of the quantum computing era poses a fundamental threat to the classical cryptographic protocols that underpin modern digital security. Algorithms like Shor's algorithm potentially compromise widely used public key systems such as RSA and ECC\cite{ref1,ref2,ref3,ref4}, thus creating an urgent need for quantum
	resistant solutions. Within this landscape, QSDC\cite{ref5,ref6,ref7,ref8,ref9} has emerged as a promising paradigm. Unlike Quantum Key Distribution (QKD)\cite{ref10,ref11,ref12,ref13,ref14,ref15,ref16,ref17}, which only secures key exchange, QSDC enables the direct transmission of confidential messages over a quantum channel without pre-sharing keys, offering a fundamental advantage.
	
	However, the practical advancement of QSDC, particularly towards multi  user scenarios requiring supervision, is constrained by a critical trilemma. This three
	way trade off involves balancing communication efficiency, protocol security, and controllability. Many existing Controlled Quantum Secure Direct Communication(CQSDC) protocols rely on complex interactions or extensive classical post processing to implement control functions. This often leads to low qudit
    efficiency\cite{ref18,ref19,ref20,ref21,ref22,ref23,ref24}, which significantly restricts their practicality in high bandwidth quantum networks. Overcoming this trilemma is a central challenge for the practical deployment of QSDC.
	
	To address this trilemma, research efforts have primarily advanced along two promising yet distinct paths. The first path explores high - dimensional (HD) quantum communication, which utilizes quantum states beyond qubits. This approach significantly enhances the information capacity per photon and offers inherent advantages in resisting eavesdropping and channel noise\cite{ref25,ref26}. Landmark experiments, such as the stable transmission over 100 km of fiber \cite{ref27} and high fidelity free space links\cite{ref28}, are rapidly accelerating its practical application. The successful implementation of a 15-user QSDC network establishes a foundational platform for large-scale deployment\cite{ref29}.
	
	Concurrently, a second parallel path has emerged through the innovative application of quantum computing algorithms to communication fundamental elements. A seminal example is the adaptation of Grover's search algorithm for QSDC protocols, demonstrating the potential of translating quantum computational advantages into enhanced communication paradigms.
	
	However, a critical examination reveals a significant fragmentation between these avenues. Research in HD encoding has often progressed with a primary focus on state preparation and transmission. Frequently, it lacks equally sophisticated, algorithm driven decoding mechanisms to fully exploit the advantages of the enlarged Hilbert space. Conversely, protocols incorporating quantum algorithms like Grover's search algorithm have largely remained confined to two dimensional systems\cite{ref19,ref20,ref22,ref23,ref24}, thereby self - limiting their potential information capacity. Most critically, a fundamental gap persists: the absence of a unified framework that organically and securely integrates HD quantum states, efficient algorithmic decoding, and a robust authentication mechanism within a CQSDC architecture. This lack of integration is the primary reason existing schemes struggle to simultaneously achieve high efficiency, provable security, and scalable control, ultimately limiting their practical applicability.
	
	This work breaks the deadlock by proposing a novel tripartite high dimensional QSDC protocol. It synergistically combines four dimensional single particle states with a reformulated Grover's algorithm, all supported by a decoy state based authentication scheme. This core theoretical innovation is the discovery and proof of a Deterministic Decoding Theorem for tripartite controlled communication. For a specific set of four dimensional initial states$|S\rangle$, the unitary sequence$ U_{S}\cdot U_{w_{C}}\cdot U_{w_{A}}\cdot U_{w_{C}}|S\rangle$ deterministically evolves it to the target message state$|w_{A}\rangle$, where $W_{A}$ is Alice's secret and $W_{C}$ is Charlie's authorization key. This theorem generalizes Grover iteration, surpassing probabilistic decoding by enabling direct message recovery without classical post processing.Architecturally, we designed an integrated protocol integrating high dimensional encoding, algorithmic decoding, and identity authentication. It leverages the high capacity of 4D states and the deterministic power of Grover operation. Also, it incorporates a decoy photon detection mechanism linked to pre - shared identity sequences for simultaneous verification of channel security and participant legitimacy. The protocol achieves a record 66.7\% qudit efficiency, outperforming existing schemes. A rigorous security analysis shows its robustness against man in the middle, intercept resend, and Trojan horse attacks.
	
	The outline of this paper is organized as follows. Section 2 briefly presents a quantum search algorithm designed for a four dimensional single particle system. Section 3 provides a detailed description of the proposed continuous variable QSDC protocol, supported by concrete examples. In Section 4, the paper conducts an in depth discussion on the security of the proposed protocol. Section 5 compares the proposed protocol with existing Quantum Secure Authentication (QSA) based protocols and evaluates their experimental feasibility. Finally, Section 6 presents a short summary conclusion. 
	
	\section{Quantum Search Algorithm for Four Dimensional Single Particle System}

    The theoretical foundation of the proposed protocol is an innovative framework that constructs a Controlled Quantum Secure Direct Communication (CSDC) scheme based on a Deterministic Decoding Theorem. Specifically, this framework integrates the core principle of Grover's algorithm into a cooperative unitary operation, enabling deterministic message retrieval upon controller authorization.	
	
	\subsection{Definition of Core Operators and the Symmetric Initial State Set}
	Grover's search algorithm enables quantum speedup through iterative application of an oracle operator ($U_{w}$) and a ​​diffusion operator​​ ($U_{s}$)\cite{ref30}. The protocol operates within a four dimensional Hilbert space $H_{4}$ ，spanned by the computational basis states  ${|0\rangle,|1\rangle,|2\rangle,|3\rangle}$. Notably, the use of a higher dimensional Hilbert space is intentional, aiming to increase the information capacity per photon and enhance robustness against channel noise. The core operators are defined as follows:
	\begin{itemize}
		\item \textbf{Oracle Operator \( (U_{w}) \)}: For a classical message \( w\in\left\{0,1,2,3\right\} \), its corresponding quantum oracle is defined as:
		
		\[U_{w}=I-2\left|w\right\rangle\left\langle w\right|\]
		
		where \( I \) is the four dimensional identity matrix.This operator is involutory$(U^{2}_{w}=I)$, and its action is to apply a selective phase flip to the basis state\( \left|w\right\rangle \).
		
		\item \textbf{Diffusion Operator \( (U_{S}) \)}: This operator performs inversion about the mean and is defined relative to the initial state \( \left|S\right\rangle \):
		
		\[U_{S}=2\left|S\right\rangle\left\langle S\right|-I\]

		\begin{table}
			\caption{The set of symmetric initial states $|S\rangle$ in the four dimensional Hilbert space for deterministic decoding}
			\begin{tabular}[htbp]{@{}lll@{}}
				\hline
				Four dimensional single particle state species \\ 
				\hline
				$S^{0}=\frac{1}{2}(|0\rangle+|1\rangle+|2\rangle+|3\rangle)$\\ 
				
				$S^{1}=\frac{1}{2}(|0\rangle-|1\rangle+|2\rangle-|3\rangle)$ \\
				
				$S^{2}=\frac{1}{2}(|0\rangle+i|1\rangle+|2\rangle+i|3\rangle)$\\ 
				
				$S^{3}=\frac{1}{2}(|0\rangle-i|1\rangle+|2\rangle-i|3\rangle)$\\ 
			
				$S^{4}=\frac{1}{2}(|0\rangle+|1\rangle-|2\rangle-|3\rangle)$ \\
				
				$S^{5}=\frac{1}{2}(|0\rangle-|1\rangle-|2\rangle+|3\rangle)$ \\
				
				$S^{6}=\frac{1}{2}(|0\rangle+i|1\rangle-|2\rangle-i|3\rangle)$ \\
				
				$S^{7}=\frac{1}{2}(|0\rangle-i|1\rangle-|2\rangle+i|3\rangle)$ \\
				
				$S^{8}=\frac{1}{2}(|0\rangle+|1\rangle+i|2\rangle+i|3\rangle)$ \\
				
				$S^{9}=\frac{1}{2}(|0\rangle-|1\rangle+i|2\rangle-i|3\rangle)$ \\
				
				$S^{10}=\frac{1}{2}(|0\rangle+i|1\rangle+i|2\rangle-|3\rangle)$ \\
				
				$S^{11}=\frac{1}{2}(|0\rangle-i|1\rangle+i|2\rangle+|3\rangle)$ \\
				
				$S^{12}=\frac{1}{2}(|0\rangle+|1\rangle-i|2\rangle-i|3\rangle)$ \\
				
				$S^{13}=\frac{1}{2}(|0\rangle-|1\rangle-i|2\rangle+i|3\rangle)$ \\
				
				$S^{14}=\frac{1}{2}(|0\rangle+i|1\rangle-i|2\rangle+|3\rangle)$ \\
				
				$S^{15}=\frac{1}{2}(|0\rangle-i|1\rangle-i|2\rangle-|3\rangle)$ \\
				\hline
			\end{tabular}
		\end{table}
	\end{itemize}
	It performs an inversion about the initial state \( \left|S\right\rangle \) in $H_{4}$. The deterministic decoding success of the protocol critically relies on the specific mathematical properties of the initial state \( \left|S\right\rangle \). To this end, we propose and formally define a set of 16 initial states, as detailed in \textbf{Table 1}. These states are not arbitrarily selected, they form a set of maximally symmetric, equiangular vectors.Crucially, their defining characteristic is that they represent uniform superpositions of computational basis states, ensuring that, prior to the oracle operator's application, the system resides in a symmetric state exhibiting a uniform probability distribution across all basis states. This symmetry is the prerequisite enabling Grover's diffusion operator $U_{S}$ to perform an exact inversion around the mean, ultimately leading to deterministic probability amplification:
	$$U_{S}\cdot U_{w}\left|S\right\rangle$$which amplifies the amplitude of the target state \( \left|w\right\rangle \), making it the most probable measurement outcome.
	
	\subsection{Deterministic Decoding Theorem for Tripartite Controlled Communication​}
	
	{\bfseries Theorem :}
	Let $w_{A}$, $w_{C} \in \{0,1,2,3\}, U_{w_{n}}=I-2|w_{n}\rangle\langle w_{n}|$ $(n=A,C)$
	, $U_{S}=2|S\rangle\langle S|-I$ be the diffusion operator for the initial state $|S\rangle$(Table1). Then, for any values of $w_{A}$, $w_{C}$:$$ U_{S}\cdot U_{w_{C}}\cdot U_{w_{A}}\cdot U_{w_{C}}|S\rangle=|w_{A}\rangle $$
	The equality holds up to an irrelevant global phase factor of $ \pm 1 $ or $ \pm i $.
	
	{\bfseries Corollary }
	For any $w_{A}$, $w_{C} \in \{0,1,2,3\}$, the composed oracle operator satisfies the identity $U_{w_{C}}\cdot U_{w_{A}}\cdot U_{w_{C}}=U_{w_{A}}$
	
	\section{Description of the proposed CQSDC protocol}
	\subsection{Executing Procedure of proposed CQSDC Protocol}
	This section provides a comprehensive description of a CQSDC protocol based on a four dimensional single particle state with Grover's algorithm. The authorized parties in the protocol are the sender (Alice), the receiver (Bob), and the controller (Charlie). We assume the receiver is honest, meaning he will faithfully execute the protocol and will not attempt to obtain or assist others in obtaining any secret information. Specifically, Alice establishes an identity authentication sequence sharing scheme with Charlie, who possesses $N$ bit authentication sequences of length $ID_{C}=\{0,1,2,3\}^{n}$. Assuming Alice aims to transmit a secret message
	$M=\{w_{A}^{j}|j=1,2,\ldots,N\}$ to Bob, where $w_{A}^{j}\in\{0,1,2,3\}$. The protocol is executed in six steps.
	
	{\bfseries Step 1:}
	Charlie generates $N$ four dimensional single particle states that functions as information carriers $S^{j}=|S^{j}\rangle$ and subsequently constructs an ordered sequence of particles $S=\{S^{j}|1\leq j\leq N\}$, where the total number of distinct four dimensional single particle state types is 16 (Table 1). At the same time, Charlie generates a random bit string $K_{c}=\{w_{c}^{j}\}_{j=1,2,\ldots,N}$ ($w_{c}^{j}\in\{0,1,2,3\}$), which serves as his license. According to the encoding rules described in \textbf{Table 2}, Charlie executes the encoding operation $U_{w_{j}^{c}}$ on the four dimensional single particle states to obtain $S_{c}=\{S^{j}_{C}|1\leq j\leq N\}$. After that, Charlie generates a decoy photon sequence $S_{IDC}$ based on his own identity sequence, where the preparation rules are outlined in \textbf{Table 3}.The corresponding X-basis state is given by:$ |k_X\rangle = \frac{1}{\sqrt{d}} \sum_{j=0}^{d-1} \omega^{j k} |j\rangle $
	,where $d=4$. Charlie then combines all the decoy photons with $S_{C}$ to form a new sequence $S^{'}_{C}$, and sends the new sequence $S^{'}_{C}$ to Alice.

	\begin{table}
		\caption{ Encoding Rules}
		\begin{tabular}[htbp]{@{}lll@{}}
			\hline
			Unitary operation $U_{i}$ & Corresponding encoding information \\
			\hline
			$U_{0}$ & 0 \\
			$U_{1}$ & 1 \\
			$U_{2}$ & 2 \\
			$U_{3}$ & 3 \\
			\hline
		\end{tabular}
	\end{table}
	
	\begin{table}
		\caption{Relationship Between Identity Sequences and Single Particle Measuring Basis}
		\begin{tabular}[htbp]{@{}lll@{}}
			\hline
			$ID_{B}$ ($ID_{C}$) & Corresponding to measuring basis \\
			\hline
			0 & $Z$ basis: $\{|0\rangle,|1\rangle,|2\rangle,|3\rangle\}$ \\
			1 & $X$ basis: $\{|X_{0}\rangle,|X_{2}\rangle,|X_{3}\rangle,|X_{4}\rangle\}$ \\
			\hline
		\end{tabular}
	\end{table}
	
    {\bfseries Step 2:}
	After Alice confirms to Charlie that she has received the quantum state sequence $S^{'}_{C}$ and completed the quantum storage, both parties initiate the joint detection procedure for quantum channel security. Firstly, Charlie synchronizes with Alice the positions of all the decoy photons and their preparation bases in the sequence $S^{'}_{C}$ via the classical channel. Based on this information, Alice extracts the corresponding decoy photons and conducts the measurements by strictly following the base vector matching rules outlined in Table 3. She then converts the measurement results into the classical bit sequence $E_{1}$ according to the rules in \textbf{Table 4} and only discloses this sequence through the authenticated classical channel.After completing the above operations, Charlie generates the local sequence $E_{C}$ based on the same rules. Subsequently, both parties jointly compute the quantum bit BER between $E_{1}$ and $E_{C}$ using the information coordination technique referenced in the literature \cite{ref31}. Only when the BER is lower than a preset threshold, the system synchronizes to achieve mutual recognition of quantum channel security authentication and participant identity legitimacy, and the protocol advances to step 3. If an abnormal BER is detected, the system immediately destroys the quantum state related to the identity mark $ID_{C}$, terminates the current session, and triggers the protocol full process reset mechanism.
	
	\begin{table}
		\caption{Classical information recording rules}
		\begin{tabular}[htbp]{@{}lll@{}}
			\hline
			Measurement results & Corresponding to classical bit \\
			\hline
			$|0\rangle/X_{0}$ & 0 \\
			$|1\rangle/X_{1}$ & 1 \\
			$|2\rangle/X_{2}$ & 2 \\
			$|3\rangle/X_{3}$ & 3 \\
			\hline
		\end{tabular}
	\end{table}
	{\bfseries Step 3:}
	After completing the security verification, Alice executes the corresponding encoding operation $U_{w^{j}_{A}}$ on the quantum states $|S^{j}_{C}\rangle$ within the sequence $S_{C}$ that she holds. She then obtains the ciphertext carrying sequence $S_{CA} =
	(S^{j}_{CA}|1\leq j\leq N)$. Alice prepares a sequence of the decoy photons $S_{D}$ ，where each decoy photon is randomly prepared by choosing from either the Z basis or the X basis. The preparation is independent of any participant's identity sequence. Alice randomly interleaves $S_{D}$ with $S_{CA}$ to form $S^{'}_{CA}$ and transmits it to Bob.
	
	{\bfseries Step 4:}
	After Bob receives and caches the sequence $S^{'}_{CA}$, he and Alice conduct security detection and authentication using the same eavesdropping detection principle as in step 2. The BER is computed by comparing Alice's error sequence $E_{2}$ with Bob's error sequence $E_{B}$. If the BER is lower than the preset threshold, the system synchronizes to accomplish mutual acknowledgment of quantum channel security authentication and participant identity legitimacy, and Bob acquires the sequence $S_{CA}$.
	
	{\bfseries Step 5:} 
	If Charlie consents to authorize this communication, he announces his license key $K_C = \{w_C^j\}$ and the initial states $|S\rangle = \{S^j | 1 \leq j \leq N\}$ to Bob via an authenticated classical channel. Subsequently, Bob executes the encoding operation $U_{w_C^j}$ on the quantum state $|S_{CA}^j\rangle$ to yield the encoded sequence $S_{CAC} = \{S_{CAC}^j | 1 \leq j \leq N\}$.
	
	{\bfseries Step 6:}
	Finally, the diffusion operator $U_{S^j}$ is applied to accomplish the Grover amplification: $|\psi_{\mathrm{final}}^j\rangle = U_{S^j} |S_{CAC}^j\rangle$, thereby obtaining the final sequence $S_{\mathrm{final}} = \{\psi_{\mathrm{final}}^j | 1 \leq j \leq N\}$. Subsequently, Bob carries out a projective measurement on each state within the sequence $S_{\mathrm{final}}$ with respect to the Z - basis. As established by the Deterministic Decoding Theorem, the measurement result for the $j$-th particle is deterministic. Specifically, it is the basis state $|w_A^j\rangle$. From this, Bob directly extracts Alice's secret message $M = \{w_A^j | j=1,2,\ldots,N\}$.
	
	\section{ Security Analysis}
	The security of the proposed CQSDC protocol is fundamentally rooted in the ​​deterministic decoding principle​​ derived from the Grover algorithm and the ​​physical laws of quantum mechanics​​. The protocol utilizes QKD to establish absolutely secure authentication sequences, ensuring the unconditional security of their generation and sharing. During execution, all authentication information remains protected by quantum encryption mechanisms. Following systematic security verification, the protocol demonstrates the capability to effectively counteract typical network attacks, including man-in-the-middle attacks and interception and replay attacks.
	
	\subsection{Controlling Parties}
	This subsection conducts a security analysis targeting dishonest controllers. The analysis demonstrates that the design of the protocol offers information theoretic security guarantees against such attacks, which are rooted in the physical laws of quantum mechanics rather than computational assumptions. The cornerstone of security lies in the combination of the principle of information carrier isolation and the irreversibility of quantum operations. In the first step, after Charlie transmits the encoded sequence $S_{C}$ to Alice, he permanently loses direct physical access to the quantum information carriers. Subsequently, the core sequence $S_{CA}$ is exchanged solely between Alice and the honest receiver, Bob. This physical isolation is not merely procedural but is cryptographically enforced by the quantum no cloning theorem, which prevents Charlie from creating an independent copy of the quantum state. Charlie's knowledge is incomplete. Although he holds his own license key $K_C$ and the initial states $|S\rangle$, he cannot access the specific quantum state $S_{CA}$ or the final sequence $S_{CAC}$ processed by Bob. Without access to these intermediate states, any attempt at reverse engineering or calculating the value of $w_{A}$ is infeasible. From Charlie's perspective, after his transmission, the system is in a maximally mixed state, making any value of $w_{A}$ equally probable and thus reducing his attack to a random guess with a negligible success probability. The protocol, relying on a robust mechanism grounded in physical laws, successfully achieves its design goal of maintaining communication confidentiality against dishonest controllers. Although it is at risk of being subjected to a Denial - of - Service (DoS) attack launched by malicious parties, this risk is within an acceptable range within the security model of controlled quantum communication. 
	
	\subsection{Man-in-the-Middle Attack}
	If the attacker, Eve, attempts to carry out a man-in-the-middle attack on the communication links between Charlie and Alice, as well as between Alice and Bob, to steal information, this protocol can effectively resist such attacks through decoy photon detection, identity authentication, and quantum state decoding control mechanisms. Taking the communication between Charlie and Alice as an example. When Eve intercepts the sequence
	$S^{'}_{C}$ sent from Charlie to Alice and impersonates Alice, the protocol starts its defense at step 2. Alice merely publicly reveals the classical bit sequence $E_{1}$, which corresponds to the measurement results of the decoy photons, while maintaining the confidentiality of the preparation basis information. Given that the basis information is not disclosed, Eve is unable to infer Charlie's identity sequence $ID_{C}$ and, consequently, cannot select the correct measurement basis to evade detection. This process not only validates the legitimacy of identities but also assesses the security of the communication channel by analyzing the quantum bit error rate (QBER). Eve's interception and subsequent resending will introduce interference, leading to abnormal QBER values and, as a result, exposing her attack.Even if Eve succeeds in passing the decoy photon detection, her attack is still ineffective. The quantum states she intercepts are encrypted with Charlie's license key $K_C$. Lacking knowledge of $K_C$, these quantum states seem entirely random and indistinguishable from a maximally mixed state to her, rendering it impossible for her to extract or tamper with the information. Even if she transmits them to Alice, the ultimate decoding authority rests with Charlie, preventing her from accomplishing the communication or acquiring the final secret information $M$.
	
	\subsection{Entangle and Measure Attack}
	In the quantum eavesdropping scenario, Eve prepares the auxiliary particle
	$e$ in the quantum state $|e\rangle$. Eve's attack operation is characterized by the unitary operation $U_{e}$. When Alice transmits the quantum sequence $S^{'}_{CA}$, which carries the secret information to Bob, Eve intercepts a subsequence of $S^{'}_{CA}$ and performs the unitary operation $U_{e}$ on the intercepted particle with the aim of extracting the secret information encoded by Alice. However, theoretical analysis indicates that this eavesdropping behavior will inevitably be detected during the security detection phase in step 4. The fundamental reason is that the coding sequence $S'_{CA}$ is intermingled with the decoy photons used for security detection. It is worth noting that after Eve performs the attack operation $U_{e}$, the states $\vert0\rangle$, $\vert1\rangle$, $\vert2\rangle$, and $\vert3\rangle$ evolve into new entangled states, respectively.
	\begin{align*}
		|\varphi_{0}\rangle &= U_{e}(|0\rangle|e\rangle) = |0\rangle|e_{00}\rangle + |1\rangle|e_{01}\rangle + |2\rangle|e_{02}\rangle + |3\rangle|e_{03}\rangle\tag{1}
	\end{align*}
	\begin{align*}
		|\varphi_{1}\rangle &= U_{e}(|1\rangle|e\rangle) = |0\rangle|e_{10}\rangle + |1\rangle|e_{11}\rangle + |2\rangle|e_{12}\rangle + |3\rangle|e_{13}\rangle\tag{2}
	\end{align*}
	\begin{align*}
		|\varphi_{2}\rangle &= U_{e}(|2\rangle|e\rangle) = |0\rangle|e_{20}\rangle + |1\rangle|e_{21}\rangle + |2\rangle|e_{22}\rangle + |3\rangle|e_{23}\rangle\tag{3}
	\end{align*}
	\begin{align*}
		|\varphi_{3}\rangle &= U_{e}(|3\rangle|e\rangle) = |0\rangle|e_{30}\rangle + |1\rangle|e_{31}\rangle + |2\rangle|e_{32}\rangle + |3\rangle|e_{33}\rangle\tag{4}
	\end{align*}
	where $|e_{xy}\rangle,(x,y)\in\{0,1,2,3\}$ is a purely auxiliary state uniquely determined by $U_{e}$, $U_{e}$ is to be an integrating operator, $U_{e}U^{+}_{e} = U^{+}_{e}U_{e} = I$,The coefficients of equations(1-4)satify the conditions such that,
	\begin{align}
		|\alpha|^{2}+|\beta|^{2}+|\gamma|^{2}+|\delta|^{2} = 1\tag{5} \\ 
		|\delta|^{2}+|\alpha|^{2}+|\beta|^{2}+|\gamma|^{2} = 1\tag{6}\\ 
		|\gamma|^{2}+|\delta|^{2}+|\alpha|^{2}+|\beta|^{2} = 1\tag{7}\\ 
		|\beta|^{2}+|\gamma|^{2}+|\delta|^{2}+|\alpha|^{2} = 1\tag{8}
	\end{align}
	If Eve attempts to eavesdrop without being detected,$\beta^{2} ,\gamma^{2} and \delta^{2}$ must be 0.In addition to this, the decoy photons include $|X_{0}\rangle, |X_{1}\rangle, |X_{2}\rangle, |X_{3}\rangle$, \\
	which, after Eve's attack, become
	\begin{align}
		|\varphi_{x_{0}}\rangle &= U_{e}(|\varphi_{x_{0}}\rangle|e\rangle) \notag \\
		&= \frac{1}{2}\big[|0\rangle(|e_{00}\rangle + |e_{10}\rangle + |e_{20}\rangle + |e_{30}\rangle) \notag \\
		&\quad + |1\rangle(|e_{01}\rangle + |e_{11}\rangle + |e_{21}\rangle + |e_{31}\rangle) \notag \\
		&\quad + |2\rangle(|e_{02}\rangle + |e_{12}\rangle + |e_{22}\rangle + |e_{32}\rangle) \notag \\
		&\quad + |3\rangle(|e_{03}\rangle + |e_{13}\rangle + |e_{23}\rangle + |e_{33}\rangle)\big] \tag{9} \\
		|\varphi_{x_{1}}\rangle &= U_{e}(|\varphi_{x_{1}}\rangle|e\rangle) \notag \\
		&= \frac{1}{2}\big[|0\rangle(|e_{00}\rangle + i|e_{10}\rangle - |e_{20}\rangle - i|e_{30}\rangle) \notag \\
		&\quad + |1\rangle(|e_{01}\rangle + i|e_{11}\rangle - |e_{21}\rangle - i|e_{31}\rangle) \notag \\
		&\quad + |2\rangle(|e_{02}\rangle + i|e_{12}\rangle - |e_{22}\rangle - i|e_{32}\rangle) \notag \\
		&\quad + |3\rangle(|e_{03}\rangle + i|e_{13}\rangle - |e_{23}\rangle -i |e_{33}\rangle)\big] \tag{10} \\
		|\varphi_{x_{2}}\rangle &= U_{e}(|\varphi_{x_{2}}\rangle|e\rangle) \notag \\
		&= \frac{1}{2}\big[|0\rangle(|e_{00}\rangle - |e_{10}\rangle + |e_{20}\rangle - |e_{30}\rangle) \notag \\
		&\quad + |1\rangle(|e_{01}\rangle - |e_{11}\rangle + |e_{21}\rangle - |e_{31}\rangle) \notag \\
		&\quad + |2\rangle(|e_{02}\rangle - |e_{12}\rangle + |e_{22}\rangle -|e_{32}\rangle) \notag \\
		&\quad + |3\rangle(|e_{03}\rangle -|e_{13}\rangle + |e_{23}\rangle -|e_{33}\rangle)\big] \tag{11} \\
		|\varphi_{x_{3}}\rangle &= U_{e}(|\varphi_{x_{3}}\rangle|e\rangle) \notag \\
		&= \frac{1}{2}\big[|0\rangle(|e_{00}\rangle -i |e_{10}\rangle -|e_{20}\rangle +i |e_{30}\rangle) \notag \\
		&\quad + |1\rangle(|e_{01}\rangle -i |e_{11}\rangle - |e_{21}\rangle +i |e_{31}\rangle) \notag \\
		&\quad + |2\rangle(|e_{02}\rangle -i|e_{12}\rangle - |e_{22}\rangle +i |e_{32}\rangle) \notag \\
		&\quad + |3\rangle(|e_{03}\rangle -i |e_{13}\rangle - |e_{23}\rangle +i |e_{33}\rangle)\big] \tag{12}
	\end{align}
	Without loss of generality, the$\langle e_{00}|e_{10}\rangle = \langle e_{00}|e_{20}\rangle = \langle e_{00}|e_{30}\rangle = \langle e_{10}|e_{20}\rangle = \langle e_{10}|e_{30}\rangle = \langle e_{20}|e_{30}\rangle = \langle e_{01}|e_{11}\rangle = \langle e_{01}|e_{21}\rangle = \langle e_{01}|e_{31}\rangle = \langle e_{11}|e_{21}\rangle = \langle e_{11}|e_{31}\rangle = \langle e_{21}|e_{31}\rangle = \langle e_{02}|e_{12}\rangle = \langle e_{02}|e_{22}\rangle = \langle e_{02}|e_{32}\rangle = \langle e_{12}|e_{22}\rangle = \langle e_{12}|e_{32}\rangle = \langle e_{22}|e_{32}\rangle = \langle e_{03}|e_{13}\rangle = \langle e_{03}|e_{23}\rangle = \langle e_{03}|e_{33}\rangle = \langle e_{13}|e_{23}\rangle = \langle e_{13}|e_{33}\rangle = \langle e_{23}|e_{33}\rangle = 0$.
	When Eve attempts undetected eavesdropping, we have $|\delta|^{2} = |\beta|^{2} = |\gamma|^{2} = 0$, $|\alpha|^{2} = 1$. As $|e'_{00}\rangle = |e'_{11}\rangle = |e'_{22}\rangle = |e'_{33}\rangle = 0$, Eve is undetectable and can't extract useful info. Thus, any attempt to get private info will introduce errors, making her entanglement measurement attack detectable during checks. The protocol resists such attacks. Beyond decoy photon detection, the protocol has an intrinsic security layer from its deterministic decoding principle. Eve's entanglement operation $U_{e}$ perturbs the quantum state, disrupting the phase relationships crucial for the Grover - based decoding sequence. Even if Eve's attack bypasses decoy photon detection, the post interaction composite state of the particle and her ancilla is $|\Psi_{e}\rangle = U_e \left( U_{w_A} U_{w_C} |S\rangle \otimes |e\rangle \right)$. Bob unaware of the attack, applies the authorized decoding sequence to the particle, not the whole system.For deterministic and correct decoding, the output must be a separable state yielding $|w_A\rangle |e'\rangle$ upon measurement. This requires $U_e$ to commute with the encoding decoding process and not entangle the particle with the ancilla. Any deviation leads to an undesired final state. The attack's ineffectiveness is quantified by fidelity $F= \left\| \langle w_A| \langle e^{'}| \cdot |\Psi_{e}\rangle \right\|^2$. Here, $F$ is the overlap between the particle's reduced density matrix in $|\Psi^{'}\rangle$ and the desired pure state, with $F < 1$. A low $F$ means a Z - basis projective measurement of the particle won't surely give $w_A$. The decoding error probability is bounded by $P = 1 - F$. So, Eve's action introduces intrinsic errors. Bob will likely fail to decode or get wrong results, providing an independent information theoretic security guarantee against entanglement attacks.

	\subsection{Intercept-and-Resend Attack}
	Eve attempts to steal information by intercepting the quantum state sequences $S^{'}_{C}$ from Charlie to Alice, and $S^{'}_{CA}$ from Alice to Bob. However, she cannot distinguish decoy photons from photons encoding information. Due to basis mismatch, any interception and measurement of the entire sequence will inevitably introduce errors in the decoy states. These errors will be detected in the form of an abnormal QBER during the public discussion phase, leading to immediate session termination. To avoid detection, Eve can only guess the measurement basis, and the probability of her
	guessing correctly is $\frac{1}{2}$. Once she guesses wrong, any of her operations will disrupt the preset deterministic process, resulting in verification failure and thus inevitable detection. Therefore, for a single decoy photon, the error probability introduced by Eve's attack is $\frac{1}{2}$. The probability that Eve passes a single decoy photon security check is also $\frac{1}{2}$. Assuming the number of decoy photons used to detect attacks is $k$, the probability that Eve's malicious behavior is discovered in each eavesdropping detection process is $P_{1}=1-(\frac{1}{2})^{k}$. We conduct a comparative analysis between our protocol and other's protocol\cite{ref18,ref19,ref20,ref22}, as shown in \textbf{Figure 1}. In protocols , if Eve guesses the measurement basis correctly, no errors are introduced. If she guesses wrong , the probability of introducing errors is $\frac{1}{2}$. Thus, for a single decoy photon, the error probability introduced by Eve's attack is $\frac{1}{4}$. The probability that Eve passes a single decoy photon security check is $\frac{3}{4}$. The probability that Eve's malicious behavior is discovered in each eavesdropping detection process is $P_{2}=1-(\frac{3}{4})^{k}$. A critical evaluation indicates that our protocol achieves a high detection probability even with a relatively low number of decoy particles $k$.
	\begin{figure}
		\includegraphics[width=0.4\linewidth]{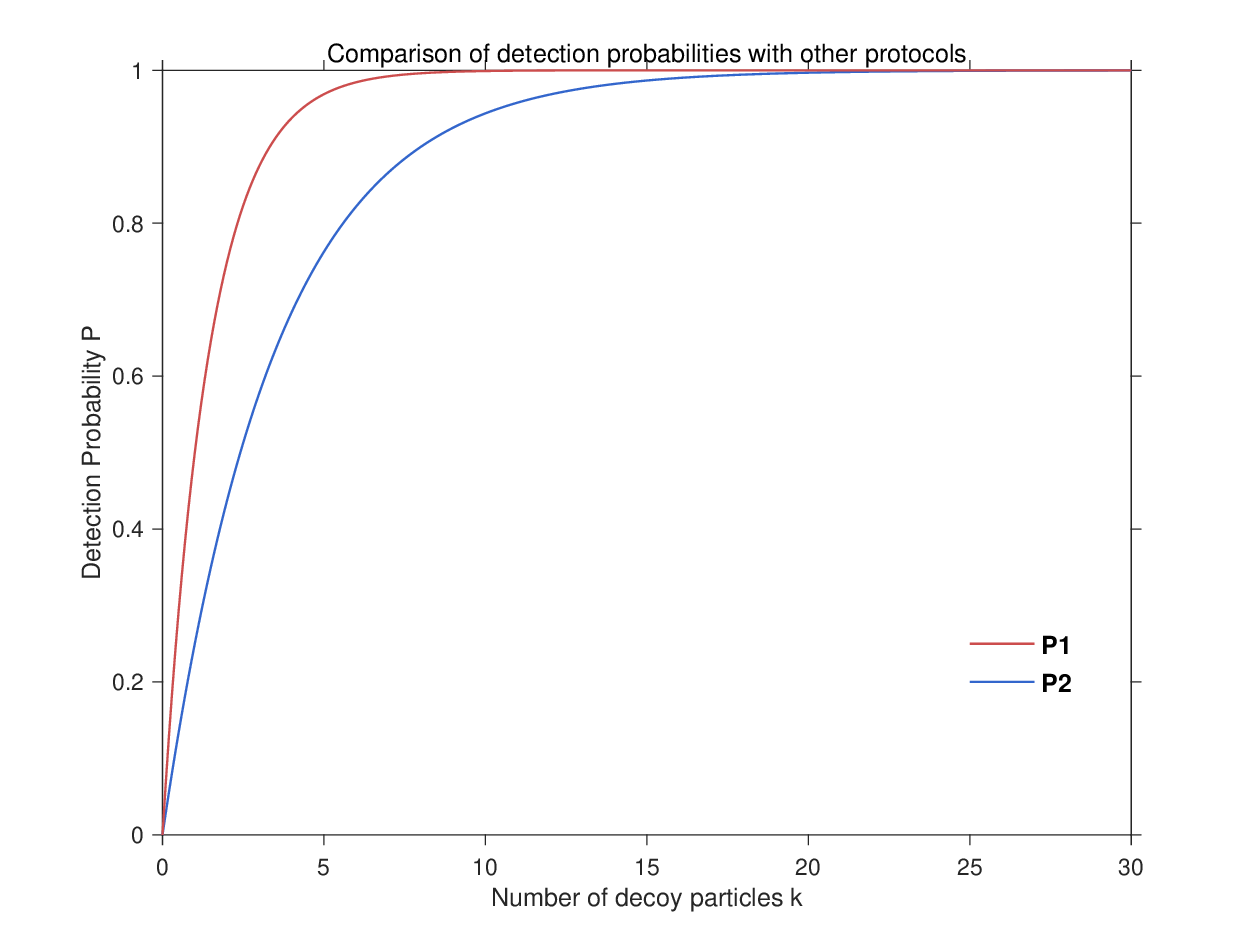}
		\caption{The red curve represents the detection probability of the proposed protocol, while the blue curve represents that of the other protocol. Both curves illustrate the relationship between the detection probability $P$ and the number of decoy particles $k$. The red curve rises more rapidly, approaching 1 even at smaller values of $k$, whereas the blue curve increases more gradually.} 
	\end{figure}
	\subsection{Trojan Horse Attack}
	An attacker launching a Trojan horse attack\cite{ref32} can penetrate the quantum channel by injecting spy photons, including invisible photons (PE)\cite{ref33} and delayed photons\cite{ref34}. This attack causes no observable perturbation in the quantum state measurement results of spoofed photons, thus no abnormal BER fluctuations during eavesdropping detection. To counter this attack, the protocol introduces a cooperative defense system with a wavelength quantum filter and a photon number separator at the quantum state preparation stage. If illegal behavior is detected during communication, Alice and Charlie immediately terminate it. In conclusion, the proposed protocol can resist Trojan horse attacks and avoid damage.
	
    \section{ Comparison and Experimental Feasibility Analysis}
	\subsection{Performance Evaluation Comparison}
	
	This section benchmarks our work against foundational QSDC protocols \cite{ref17} and current CQSDC approaches \cite{ref19,ref20,ref23,ref35}, with a focus on architectural design, efficiency and security.Qudit efficiency is defined as per \cite{ref20,ref36}:
	\begin{equation}
		\eta = \frac{b_{s}}{q_{t}}\notag
	\end{equation}
	
	Here, $b_{s}$ and $q_{t}$ denote the total number of secret messages transmitted and the number of photons generated for quantum transmission and security detection in the protocol, respectively. Throughout the communication process of the protocol, the system consumes \( q_{t} = 3N \) quantum bits. Charlie prepares \( N \) four dimensional single particle states and \( N \) detection particles as core quantum resources, while Alice simultaneously generates \( N \) detection particles for security verification. Valid secret information of \( 2N \) quantum bits is transmitted from Alice to Bob.The protocol achieves an overall quantum efficiency of \( 66.67\% \), significantly outperforming existing CQSDC schemes. Such as 18.2\%​​ for Tseng et al \cite{ref19}, ​​20.0\%​​ for Kao with Huang \cite{ref20}, and approximately ​​25\%​​ for Yang et al \cite{ref23}.This improvement is mainly due to the doubled information capacity per photon from four dimensional encoding and the elimination of classical post processing via deterministic decoding.Compared to the traditional QSA - based QSDC \cite{ref18}, which uses two dimensional states and probabilistic decoding, our protocol adopts a four dimensional single particle system and employs the Grover iteration for deterministic decoding. This shift from probabilistic to deterministic state recovery fundamentally changes the resource overhead and security model. The authentication method shows an integrative approach. While protocols like \cite{ref19} and \cite{ref22} use decoy states for channel monitoring, our scheme simultaneously validates channel security and participant identity by binding decoy photon preparation to a pre - shared identity sequence, a feature not covered in the compared works. Four dimensional states also offer a comparative advantage in resilience to channel noise due to the larger Hilbert space, a property not utilized by two dimensional counterparts \cite{ref18,ref19,ref20,ref22,ref35}. A detailed performance comparison is presented in \textbf{Table 5}.

	\begin{sidewaystable}
		\centering 
		\caption{Comparison of existing QSA based protocols with this protocol}\label{tab5}
		\scriptsize
		\setlength{\tabcolsep}{4pt}

		\begin{threeparttable}
			\begin{tabularx}{\linewidth}{@{}l*{6}{>{\centering\arraybackslash}X}@{}}
				\hline
				& Wang et al.'s protocol\cite{ref18} 
				& Tseng et al.'s protocol\cite{ref19} 
				& Kao et al.'s protocol\cite{ref20} 
				& Yang et al.'s protocol\cite{ref22} 
				& Yin et al.'s protocol\cite{ref35} 
				& Proposed agreement \\
				\hline
				Controlling party & No & Yes & Yes & Yes & Yes & Yes \\
				Quantum efficiency  & 33.3\% & 18.2\% & 20\% & 25\% & 25\% & 66.7\% \\
				Quantum source & Direct product state  & Direct product state  & GHZ state & Product state & Beel state & Four-dimensional single particle state  \\
				Security detection strategy & Entrap a photon & Entrap a photon & Entrap a photon & Decoy photons and identity sequences & To decoy photons & four-dimensional decoy photon and identity sequences \\
				Resistance to intermediary attacks & No & No & No & Yes & No & Yes \\
				Resisting attacks by illegal receivers & No & No & No & Yes & No & Yes \\
				Resisting Trojan Horse Attacks & No & Yes & Yes & Yes & Yes & Yes \\
				\hline
			\end{tabularx}
		
			\begin{tablenotes}
				\item[(a)] ``Yes'' means that the protocol satisfies the item on the left of the line, while ``No'' means not.
			\end{tablenotes}
		\end{threeparttable} 
	\end{sidewaystable}
	\newpage
	
	\subsection{Experimental Feasibility Analysis}
	
	The practical implementation of our proposed CQSDC protocol is strongly supported by rapid advancements in several key quantum technologies. Recent experimental breakthroughs have provided a clear roadmap for implementing this protocol. They have transformed it from an elegant theoretical concept into a tangible goal for the quantum communication community. The protocol's reliance on the generation, manipulation, and measurement of four dimensional single particle states aligns perfectly with the cutting edge of photonic quantum encoding, especially in spatial mode and time bin processing. The deterministic preparation of such states is no longer a fundamental challenge but an engineering one, achievable through established methods like photonic integrated circuits \cite{ref37} or spontaneous parametric down conversion (SPDC) with custom made phase masks\cite{ref38}.Recent work, such as \cite{ref39}, provides compelling evidence for the robustness of high dimensional encoding under real world conditions. It demonstrated high fidelity transmission of complex spatial mode entangled states over kilometer scale fibers. This advancement shifts the focus from mere transmission reliability to the efficient extraction of information embedded in high dimensional states. To this end, our work adapts Grover's algorithm to perform deterministic decoding, which enables accurate information retrieval, on single 4D states via the unitary operations $U_{w}$ and $U_{S}$. These operations act as tailored phase and diffusion gates specifically designed for a single qudit system. Their experimental implementation is readily achievable using programmable linear optical elements, such as multi plane light conversion (MPLC) for spatial modes or dynamic waveplates for time bin qudits \cite{ref40}. Furthermore, the protocol's security layer, which combines decoy state photon detection with identity sequence authentication, leverages a well established technique from QKD. Integrating this security layer into our scheme incurs negligible experimental overhead. Beam splitters for sampling, high efficiency single photon detectors, and fast active switches, which are all commercially available, have been seamlessly integrated into field deployed QKD systems, demonstrating their maturity and reliability. We are confident that a proof of principle demonstration of this protocol is highly feasible with current technology in a controlled laboratory setting, leveraging existing infrastructures for high dimensional quantum communication.
	
	\section{ Summarize}
	In this work, we have proposed and thoroughly analyzed a novel CQSDC protocol that successfully integrates high dimensional quantum states, Grover's search algorithm, and a robust identity authentication mechanism into a unified security framework. By utilizing four dimensional single particle states as information carriers, we have inherently increased the channel capacity and noise resilience compared to conventional two dimensional protocols. The innovative application of Grover's algorithm, through the $U_{w}$ and $U_{S}$ soperators, enables ​​deterministic decoding​​. This is a significant advancement over probabilistic methods, eliminating the need for classical post processing and boosting the qudit efficiency to 66.7\%. Furthermore, the protocol embeds security at its core by seamlessly weaving decoy state based quantum channel verification with participant identity authentication, creating a multi layered defense that rigorously resists a comprehensive suite of quantum attacks, including man-in-the-middle, intercept resend, and Trojan horse
	attacks. It is worth noting that the identifier authenticated based on quantum key distribution technology has recyclable characteristics \cite{ref41}. This work demonstrates that algorithmic quantum advantages, typically explored in computational contexts, can be powerfully repurposed to enhance communication security. It provides a scalable blueprint for ​​multi party quantum communication networks​​ where controlled, secure information transfer is paramount. The protocol's design, which mandates controller authorization for successful decoding, establishes a verifiable and physically enforced access control mechanism, making it highly suitable for scenarios requiring hierarchical security clearance. Looking further ahead, extending this framework to ​​multi controller settings​​ and exploring its integration with ​​quantum repeaters​​ and ​​quantum memories​​ for long distance communication represent exciting challenges that will further solidify the role of CQSDC in the future quantum internet.
	
	\bibliographystyle{MSP} 
	\bibliography{references} 
\end{document}